\documentclass{emulateapj}  

\newcommand{\rL}{{\ensuremath{r_{\rm L}}}}

\begin{document}

\title{The Origin of Dark Matter Halo Profiles}
\author{Neal Dalal}
\affiliation{CITA, University of Toronto, 60 St.\ George St., Toronto,
  Ontario, M5S 3H8, Canada}
\author{Yoram Lithwick}
\affiliation{Department of Physics and Astronomy, Northwestern
  University, 2145 Sheridan Rd., Evanston, IL 60208} 
\author{Michael Kuhlen}
\affiliation{Theoretical Astrophysics Center, University of California
  Berkeley, Berkeley, CA 94720}

\begin{abstract}
A longstanding puzzle of fundamental importance in modern cosmology
has been the origin of the nearly universal density profiles of dark
matter halos found in N-body simulations -- the so-called NFW profile.
We show how this behavior may be understood, simply, by applying
adiabatic contraction to peaks of Gaussian random fields.  We argue
that dynamical friction acts to reduce enormously the effect of random
scatter in the properties of initial peaks, providing a key
simplification.  We compare our model predictions with
results of the ultra-high resolution Via Lactea-II N-body
simulation, and find superb agreement.  We show how our model may be
used to predict the distribution of halo properties like concentration.
Our results suggest that many of the basic properties of halo
structure may be understood using extremely simple physics.
\end{abstract}
\keywords{}

\section{Introduction}

Hierarchical structure formation is a messy process.  The
assembly of virialized objects in Cold Dark Matter (CDM) cosmologies 
involves countless merger and accretion events that can convulse
the interiors of growing halos.  N-body simulations have shown
that halos can have tumultuous lives, in which they are bombarded
on all sides by infalling clumps of material. 

Out of this seeming chaos, however, emerges remarkable regularity.
Simulated dark matter halos have fairly generic radial 
profiles, with density scaling as $\rho\propto r^{-3}$ at large radii,
becoming more shallow at smaller radii, approaching $\rho\sim r^{-1}$
near the resolution limit of the simulations. 
\citet[hereafter NFW]{NFW96,NFW97} 
suggested that this behavior is universal among CDM
halos.  Subsequent numerical work found qualitatively similar results, 
producing halos whose density rises steeply down to the resolution
limits of the simulations 
\citep[e.g.][]{Moore98,ViaLactea1,Gao08,GHalo,ViaLactea2,Aquarius}.
The vast majority of halos show this behavior; the exceptions largely
appear to correspond to `bridged' halos artificially linked together,
or halos that have undergone recent major mergers and not yet had time
to virialize \citep{Lukic09}.

The origin of this near-universality of halo structure has been a
longstanding puzzle that has attracted considerable theoretical
attention.  Many different types of arguments and mechanisms have been
advanced to account for halos' NFW-like behavior.  
For example, \citet{NusserSheth99} suggested that the shape of
the halo profile may be related to the shape of the matter power
spectrum.  Other groups have argued that tidal disruption
of substructure in halos could dynamically drive the inner profile
towards $r^{-1}$ \citep{SyerWhite98,Dekel03a,Dekel03b}.  While such
mechanisms could affect halo profile shapes, however, these mechanisms
are not apparently {\em required} 
to produce NFW-like profiles.  In particular, calculations of 
monolithic collapse of halos has shown that the resulting halos
generically have NFW-like profiles \citep{Huss99}.  More recently,
\citet{WangWhite09} used simulations of Hot Dark Matter cosmologies to
show that halos forming at the cutoff scale of the power spectrum have
radial profiles that are fit by the NFW form just as well as CDM halos
are at comparable resolution.  
These halos have no subhalos in them, and the power spectrum
shape on the relevant scales is completely different than CDM power
spectra, and yet the same generic halo profile results.

Because such a broad class of initial perturbations produce collapsed
halos with NFW-like profiles, it is useful to study particularly
simple classes of collapsing halos to help identify the important
mechanisms.  Towards this end, in a companion paper 
\citep[hereafter Paper I]{LD10} we studied the collapse of initially
scale-free density profiles.  This problem admits a similarity
solution, which allows us to achieve considerably higher resolution
than is possible using conventional N-body techniques.  We showed in
Paper I that NFW-like profiles are not a generic outcome of cold,
dissipationless gravitational collapse in three dimensions.  Instead,
self-similar collapse produces halos with central cusps whose shapes
depend on the slopes of the initial peaks of the linear density.  We
identified adiabatic contraction as a crucial mechanism in setting the
overall shape of the halo, and showed that the conserved adiabatic
invariants governing the contraction of mass shells are set during the
initial, quasi-linear regime preceding collapse into the halo.  Using
this result, we were able to write down a simple toy model to predict
the final collapsed profiles of halos as a function of the initial
peaks.  In this paper, we will attempt to generalize and apply these
results to the formation of halos in the context of hierarchical
structure formation.

This paper is structured as follows.  In \S\ref{sec:collapse}, we
briefly review the results of our study of self-similar collapse from
Paper I, and discuss how these results may be generalized to peaks
that are not scale-invariant.  In \S\ref{sec:VL}, we compare our model
predictions with the high-resolution N-body simulation 
Via Lactea-II.  We show that our model agrees well with the
simulation's results, however we find evidence that dynamical friction
significantly modifies the structure of the collapsed halo.  In
\S\ref{sec:stats} we present a simple model for how to account for the
effect of dynamical friction on the statistics of the hierarchy of
peaks within peaks of Gaussian random fields, again finding excellent
agreement with the Via Lactea-II simulation.  We show how our model
may be used to predict the distribution of halo properties like
concentration in \S\ref{sec:conc}, and conclude in
\S\ref{sec:conclude}.

\section{Halo Collapse} \label{sec:collapse}

Many aspects of halo formation may be understood from fairly general
considerations of gravitational collapse, as we will show.  We start
by reviewing spherical collapse of peaks.  
It will prove useful throughout the discussion in this section to
imagine decomposing the initial volume around the peak into Lagrangian
shells, indexed by their Lagrangian radius $\rL$, and to build up the
collapsed halo profile by summing over Lagrangian shells.  Our
discussion and notation here intentionally parallels that in Paper I,
however in this paper we will use conventional mass and distance
coordinates, rather than the self-similar units of Paper I.  Besides
Paper I, there have been many previous papers employing a similar
approach towards understanding halo formation; see e.g.\ 
\citet{Peebles80,RydenGunn87,White92,DelPopolo09} for examples.

Consider a spherically symmetric peak of the initial linear density field 
${\bar\delta}_{\rm lin}(\rL)$.  Here, ${\bar\delta}$ refers to the
average interior mass overdensity $\delta M/M$, not the local
overdensity $\delta\rho/\rho$, and $\rL$ is Lagrangian radius.
The spherical collapse model \citep{GunnGott72} shows that turnaround
occurs at the scale where the linear mass overdensity becomes of order
unity, that is ${\bar\delta}_{\rm lin}(r_c)\approx 1$.  The linear
density grows like the linear growth factor, $D(a)\approx a$ at early
times, so we can easily determine the time at which different scales
turn around and collapse, given the linear density profile.  For
example, if the linear density has some local slope $\gamma$, such
that locally ${\bar\delta}_{\rm lin}\propto \rL^{-\gamma}$, then 
the expansion factor $a$ when a given scale $\rL$ collapses
behaves as $a\propto \rL^\gamma$.  Therefore the turnaround radius
in proper (not comoving) coordinates behaves as $r\propto
\rL^{1+\gamma}$.  From the dependence of the turnaround radius with
initial Lagrangian radius, we can infer the halo profile under
various assumptions.

\subsection{Apoapses}
\label{sec:frozen}

First, for simplicity, let us assume (briefly) that subsequent to
turnaround, matter shells remain fixed at some constant fraction of
the turnaround radius.  We refer to this as the ``frozen'' model,
since instead of allowing mass elements to orbit, we freeze them at
their turnaround radii.  This is essentially identical to the
``circular orbit'' model of \citet{RydenGunn87}, who froze mass shells
at one-half the turnaround radius. The point of this exercise is to 
isolate the effect that the initial peak profile has on the
distribution of orbital apoapses, before shell crossing occurs.  This
is important in setting the outer profile, though at smaller
radii the orbital motions of particles lead to significant changes in
the mass distribution.

\begin{figure}
\centerline{\includegraphics[width=0.4\textwidth]{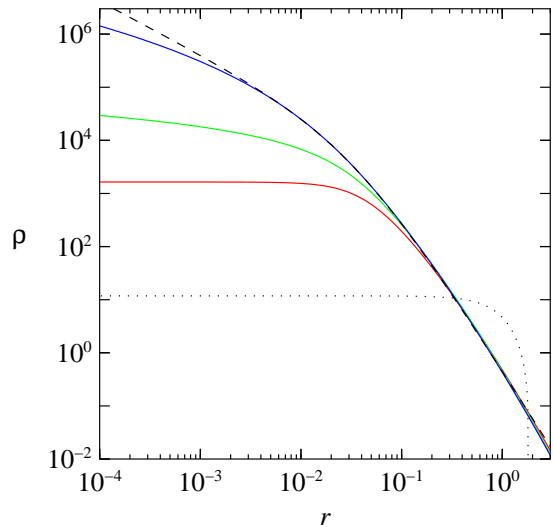}}
\caption{Simplest 1-D models for the halo profile.  The dotted black
  line shows the input linear overdensity profile, which was adapted from
  the stacked Lagrangian profiles of low-mass halos from the N-body
  simulations of \citet{Dalal08}.  It has vanishing slope at small
  radius, and a divergent slope at finite radius due to the overdensity
  becoming negative.  The solid curves show the predicted profile for
  various 1-D models.  The red curve shows the prediction of
  the frozen model, as time $t\to\infty$.  The flat region
  in the linear profile leads to the flat region in the halo profile,
  while the steep region in the linear profile produces the roughly
  $r^{-3}$ slope in the outer halo.  The green curve shows the profile
  if each shell lays down interior profiles with minimal tails,
  without any resulting adiabatic contraction.  The density is larger
  than the frozen model prediction, but still rolls over in slope
  quickly. The blue curve shows the
  prediction of the minimal contraction model.  Once any contraction
  is taken into account, the roll-over in slope from steep to shallow
  becomes considerably more gradual with radius than the frozen model.
  For comparison, the black dashed curve shows the NFW profile,
  normalized to match the minimal contraction model at $r_{-2}$, the
  radius where the logarithmic slope $d\log\rho/d\log r=-2$.
  At small radius, the slope of the minimally contracted profile rolls
  over more quickly than the NFW slope. Note that for all the 1-D
  models, the actual outer slope is slightly steeper than -3, since
  the total mass of the halo is finite. 
  \label{fig:frozen}
}
\end{figure}

Making this `frozen' assumption, we can easily determine the density
profile using $\rho\propto d^3\rL/d^3r$.  For example, if the linear
density is a power-law with slope $\gamma$, then as noted above
$r\propto\rL^{1+\gamma}$, and the density becomes 
$\rho\propto (\rL/r)^2d\rL/dr\propto r^{-g}$, 
where $g=3\gamma/(1+\gamma)$ is the
Fillmore-Goldreich slope described in Paper I.  Figure \ref{fig:frozen}
illustrates this behavior.  In regimes where the slope of the linear
density is very shallow, $\gamma\sim 0$, then the frozen model
predicts that the final collapsed profile is also shallow, 
$g\sim 0$.   In contrast, when the initial slope is quite steep,
$\gamma\to\infty$, then the final slope also becomes steep,
$g\sim 3$ (actually slightly steeper when ${\bar\delta}$ is not an
exact power law, i.e.\ when $d\gamma/dr$ is nonzero). 

This already explains some of the behavior found in halos in
cosmological N-body simulations, which tend to have steep $\sim
r^{-3}$ outer profiles.  As can be seen from the simple argument
above, this is a natural outcome whenever the initial linear density
peaks acquire steep outer slopes (e.g. when the linear overdensity
falls below zero, a common occurrence among peaks of random fields).
Put in more physical terms, steep $r^{-3}$ 
profiles naturally occur whenever the growth of halos slows or stops.
Indeed, once the halo stops growing, then as time progresses the
$r^{-3}$ region simply extends.  Essentially the same argument has
previously been used to explain the correlations between halo
concentration and formation time, which underlies the
mass-concentration relation at low mass
\citep{Wechsler02,Zhao03a,Zhao03b,Lu06}. We discuss this point in more
detail in \S\ref{sec:conc}.

\subsection{Periapses and shell profiles}

The frozen model described above assumes that each Lagrangian shell
lays down a thin annulus of material in the collapsed halo, with no
shell crossing.  This aspect is unrealistic, of course -- following
collapse, material entering the halo will execute orbits that cover a
finite range of radii, and therefore the density profile laid down by
that material is not a thin annulus, $\rho\neq\delta(r-r_{\rm apo})$,
but instead has tails extending to small radii. This
affects the overall halo profile in multiple ways.  First, the slope
of the interior profile will obviously be steeper than the
Fillmore-Goldreich slope if the tails from individual shells are steep
enough.  Secondly, because of shell crossing, the mass interior to a
Lagrangian shell grows following collapse, which causes contraction
and hence steepening of the overall profile.  

A full description of these effects requires numerical calculation
(see Paper I), however we can understand some of the basic behavior 
using simple qualitative arguments.  We can generally expect the
density profile $\rho_{\rm shell}(r)$ laid down by some shell with
total mass $M_{\rm shell}$ to take the form 
\begin{equation}
\rho_{\rm shell}(r) \propto \frac{M_{\rm shell}}{r^3}\times
\frac{t(r)}{t_{\rm orb}}\times P(r_{\rm peri}<r).
\end{equation}
Each term in this expression is easy to understand.  First, we expect
the density at each scale to behave like mass divided by volume,
i.e.\ $M_{\rm shell}/r^3$.  The middle term corresponds to the fraction of the
orbital time that a particle spends inside radius $r$, assuming it
reaches radius $r$.  Lastly, the third
piece is simply the probability that the particle is on an orbit whose
periapse is inside radius $r$; particles obviously deposit no mass at
radii $r<r_{\rm peri}$ inside of their periapses.  

Next, the time spent at radius $r$ scales as $r/v$, where $v$ is
the velocity.  Now, as long as the circular velocity profile is either
flat or declining towards small radius (i.e. the total density is 
$\rho\sim r^{-2}$ or shallower), then the velocity of a particle at
radii smaller than apoapse will be approximately constant, so that
$t(r)\propto r$.  Therefore, to a reasonable approximation, the
density laid down by a particle behaves as $\rho \propto r^{-2}$ for
$r_{\rm peri} < r < r_{\rm apo}$.  There are small corrections to this
near periapse and apoapse, but these corrections are unimportant when
we sum over all particles.   Accordingly, the density from a shell
behaves like $\rho_{\rm shell}(r) \propto r^{-2} P(r_{\rm peri}<r)$,
i.e. the density profile depends upon the distribution of periapses.

For example, in spherical collapse where all motion is perfectly
radial, all particles pass through the origin, meaning that 
$P(r_{\rm peri}<r)=1$ for all finite $r$.  Then we expect the interior
profile from each shell to behave as $r^{-2}$, and as long as the
Fillmore-Goldreich slope is shallower than this (i.e.\ $\gamma<2$),
then the interior profile will be $r^{-2}$.  This is indeed the
behavior found in the full solutions to spherical collapse
\citep{FG84,Bert85}, as described in Paper I.

More generally, however, the orbits of particles in the collapsed
halo are not perfectly radial, meaning that the interior profile is
asymptotically shallower than $r^{-2}$.  We cannot quantitatively
determine the periapse distribution without characterizing the full
orbital structure, which requires numerical calculation.  However we
can make useful qualitative arguments.  Assuming that
the density profile is shallower than $r^{-2}$, then particles at
small radius $r\ll r_{\rm apo}$ have velocities much greater than the
local circular velocity.  We can therefore treat their motion as
unaccelerated, i.e.\ straight lines.  The distribution of the
orientations of these straight lines then determines the periapse
distribution.  For example, for random orientations with some spread
(e.g.\ a Gaussian distribution), the probability
to reach radius $r$ simply scales like the area subtended by
radius $r$, that is $P(r_{\rm peri}<r)\propto r^2$.  In this case, 
we would expect that at asymptotically small
radius, the density profile laid down by individual shells should tend
towards a constant, $\rho_{\rm shell}\to\,$const as $r\to 0$.  Note
that this disagrees with \citet{Lu06}, who claimed that an isotropic
velocity dispersion leads to a $r^{-1}$ density profile, due to an
unphysical ansatz they assumed.

\subsection{Shell crossing and contraction}

When material is accreted onto halos, its mass is deposited over a
range of radii.  This newly deposited mass not only adds to the
existing density, but also causes contraction of the mass already
present in the halos.  To see this, note that the radial action
$J_r\equiv\oint v_r dr \propto [r\times M(r)]^{1/2}$ is an adiabatic
invariant for spherically symmetric systems.  If $M$ at radius $r$
increases due to accretion of new matter, then the orbits of particles
at that radius will shrink in order to hold fixed the product $r\times
M(r)$.  This effect is caused entirely by the density tails described
in the previous subsection; in the absence of such tails (i.e., in the
absence of shell crossing) the mass interior to any shell would remain
constant and there would be no further contraction following collapse.

Because the contraction acts to hold fixed $r\times M$, we can easily
estimate its effects given the $r\times M$ profile before shell
crossing (e.g.\ at turnaround) and the density tails laid down
by each shell.  The most conservative estimate of this contraction
arises if we assume minimal tails.  As noted above, in triaxial
potentials, we expect shell profiles to behave as $\rho\to\,$const as
$r\to0$, so the weakest tails we expect would be for perfectly
constant density $\rho=\,$const for all $r<r_{\rm apo}$.  We denote
this the minimal contraction model.

In this minimal contraction model, a Lagrangian mass shell 
$M_{\rm L}$, of width $dM_{\rm L}$ and apoapse $r_{\rm apo}(M_{\rm L})$ 
lays down a mass profile $M(r)=dM_{\rm L} f(r/r_{\rm apo})$, where
\begin{equation}
f(r/r_{\rm apo})=
\cases{
\left(\frac{r}{r_{\rm apo}}\right)^3, & $r<r_{\rm apo}$ \cr
1, & $r>r_{\rm apo}$ \cr
}.
\label{eqn:mininmalshell}
\end{equation}
We compute the total mass profile by summing over all shells.
This requires knowing $r_{\rm apo}(M_{\rm L})$, which we can estimate
by assuming that the product $r\times M$ is constant.  At turnaround,
this product is 
\begin{equation}
F(M_{\rm L})\equiv M_{\rm L} \times r_{\rm ta}(M_{\rm L}) \propto 
M_{\rm L}^{4/3} / {\bar\delta}_{\rm lin}(M_{\rm L}),
\label{eqn:Mr}
\end{equation}
where we have used $r_{\rm ta} \propto \rL /{\bar\delta}_{\rm lin}$. 

Let us write $r_{\rm apo}(M_{\rm L})$ as the apoapse for shell 
$M_{\rm L}$, and the inverse function $M_a(r)$ which gives the
Lagrangian shell $M_{\rm L}$ whose $r_{\rm apo}=r$.  Then we can
easily write down the total (contracted) mass profile at $t=\infty$ as
a sum over all shells \citep[e.g.][]{RydenGunn87,Lu06}
\begin{equation}
M(r) = \int dM_{\rm L} \, f\left(\frac{r}{r_{\rm apo}(M_{\rm L})}\right).
\end{equation}
Inserting our expression for the shell profile,
Eqn.\ (\ref{eqn:mininmalshell}), gives
\begin{equation}
M(r) = M_a(r) + \int_{M_a(r)}^\infty dM_{\rm L}
\left(\frac{r}{r_{\rm apo}(M_{\rm L})}\right)^3.
\end{equation}
Differentiating this expression with respect to $r$ gives
\begin{eqnarray}
\frac{dM}{dr} &=& \frac{3}{r} [M(r)-M_a(r)] \nonumber\\
&=& \frac{3}{r} [M - F^{-1}(M\,r)], 
\label{eqn:min}
\end{eqnarray}
where $F^{-1}$ is the inverse function of the expression in
Eqn.~(\ref{eqn:Mr}). This is a simple ordinary differential equation
for $M(r)$ which may easily be solved for any linear density profile
$\delta_{\rm lin}(\rL)$.  

As the name implies, the minimal contraction model provides a lower
limit on the effects of adiabatic contraction.  If the shell profiles
have more mass at small radius than this model assumes, then
contraction can be considerably stronger.  To illustrate this, we
provide another example toy model, (very) loosely motivated by our
earlier study of self-similar triaxial collapse.  We found that in
certain cases, the shell profiles had inner slopes
$d(\log\rho_{\rm shell})/d(\log r)\sim \frac{1}{2}
d(\log\rho_{\rm tot})/d(\log r)$.  If we assume that all shells have
such profiles inside of their apoapses, then following the same
reasoning that led us to Eqn.\ (\ref{eqn:min}), it is straightforward
to show that the mass profile satisfies
\begin{equation}
\frac{dM}{dr} = \frac{3M}{r} 
\frac{M - F^{-1}(M\,r)}{M + F^{-1}(M\,r)}.
\label{toy2}
\end{equation}
We label this as the $\rho^{1/2}$ model.  We stress that both of
these toy models are not meant to be considered as serious
descriptions of the shell profiles; rather they are meant to be 
illustrative, since they allow the halo mass distribution to be
computed by solving simple ordinary differential equations.

Figure \ref{fig:frozen} illustrates the effects of contraction.  The
predicted halo profile is significantly modified compared to the
frozen model, even with minimal contraction.  
As the figure shows, the transition from steep to
shallow slopes becomes much more gradual, once the effects of
contraction are taken into account.  Even though the input peak
profile sharply transitions to a flat, $\delta\sim\,$const behavior, 
similar to a top-hat perturbation, the adiabatically contracted halo
profile is very similar in form to NFW, even with minimal contraction.  
It is therefore not entirely surprising that NFW-like profiles arise
in contexts like monolithic collapse or simulations with truncated
power spectra.  

It is worth noting, however, that our models do not generically
predict power-law central cusps in the final halo profile $\rho(r)$.
Only for power-law initial profiles, ${\bar\delta}\propto r^{-\gamma}$
do we find power-law cusps, $\rho\propto r^{-g}$ (see Paper I).  More
generally, for peaks with ${\bar\delta}\to{\rm const}$ as $\rL\to 0$,
our models predict only logarithmic divergence of the halo density
$\rho$ as $r\to 0$.  The logarithmic slope $d\log\rho/d\log r\to 0$ as
$r$ approaches 0, however the roll-over in slope occurs extremely
slowly over many decades in radius.  Recent high resolution
simulations \citep[e.g][]{GHalo,Aquarius} have found similar behavior,
in the sense that their halo profiles appear better described with
rolling profiles like the Einasto profile \citep{Merritt05} instead of
power-law cusps. We return to this topic in \S\ref{sec:conclude}.

To summarize this section, 
we have described a simple method to translate from the 
initial peak to the final halo, essentially by adiabatically
contracting the linear density profile and adopting a prescription for
the distribution of orbits.  In the next section, we compare this
model to results of a high resolution cosmological N-body simulation.

\section{Comparison with N-body simulations}
\label{sec:VL}

The model discussed in the previous section may seem overly
simplistic.  It employs arguments based on spherical symmetry, and
describes halo assembly as an orderly process, resulting in an
effectively stratified structure in which the orbits of particles
within the halo reflect the locations where those particles
originated.  In contrast, the assembly of halos observed in
cosmological simulations is often violent, involving
discrete, stochastic accretion events that frequently take the form 
of major mergers.  For this reason, it is unclear whether a model that
assumes that all halo material evolves adiabatically is capable of
describing the messy reality of structure formation.

Previous work does, however, give us reason to hope that our simple
model may nonetheless be useful.  Results from multiple simulations
have shown that hierarchical assembly, including major mergers,
does not completely obliterate all pre-existing structure within halos 
\citep[e.g.][]{Kazantzidis06,Valluri07}.  More recently,
\citet{Wang10} analyzed high-resolution simulations of individual
halos, and showed that their halo structure is indeed somewhat
stratified, with a clear gradient of accretion time with radius.  On
the other hand, they also showed that major mergers can disrupt this
stratification, by bringing in fresh material into the halo core.
Therefore, it is not clear whether our model, with its simplistic
assumptions, would be relevant for realistic halos.

To test the assumptions and predictions of our simple model, we have
analyzed results of the ultra-high resolution Via Lactea-II
(VL2) simulation \citep{ViaLactea2}.  
This simulation resolves the Lagrangian region of a
typical Milky Way dark matter halo with just over one billion
particles of mass 4,100 $M_\odot$, 
and follows its evolution and formation in a
cosmological environment (40 Mpc), most of which is covered only
with lower resolution (higher mass) particles. The simulation begins at
redshift $z=104.3$, and outputs 400 snapshots in time ending at $z=0$.
The snapshots were processed to determine halo properties including
mass, centroid, and mean velocity, allowing us to reconstruct the
halo's detailed assembly history.  At $z=0$ the main
halo has $r_{200}=402$ kpc (the radius enclosing a density of $200
\rho_{M} = 200 \Omega_M \rho_{\rm crit}$) and $M_{200}=1.92
\times 10^{12} M_\odot$. Its density
profile is well fit by an Einasto profile with parameters
$\rho_{-2}=9.91 \times 10^5 M_\odot$ kpc$^{-3}$, $r_{-2}$ = 28.9
kpc, and $\alpha=0.142$.  Assuming that this simulation provides a
typical example of halo formation for galaxies like the Milky Way, its
high resolution should allow us to test the key assumptions and
predictions of our model.

The most important assumption of our model that we would like to test
is adiabaticity of the orbits.  A direct check of this for over 1
billion particles with 400 snapshots would be a computational
challenge, which we defer to future work.  Instead, we will attempt an
easier exercise which worked well in our study of self-similar
triaxial collapse in Paper I.  As mentioned in \S\ref{sec:collapse},
we expect that the product $r\times M(r)$ should be approximately
conserved if the radial action $J_r=\oint v_r dr$ is an adiabatic
invariant.  In Paper I, we computed the average of this quantity,
${\bar r}\times M({\bar r})$ for entire shells of particles, and
showed that the shell average is not only well conserved, but
furthermore may be predicted from the linear density profile of the
initial peak.  Because this is much simpler to compute, we perform a
similar test for the VL2 halo.

\begin{figure}
\centerline{\includegraphics[width=0.45\textwidth]{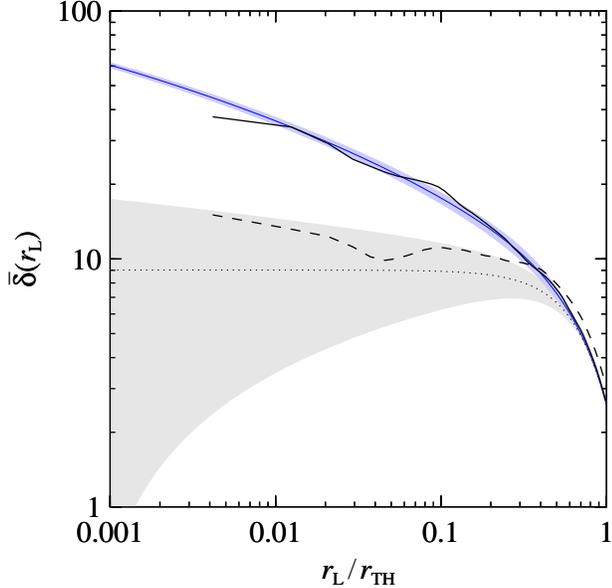}}
\caption{Initial linear density profile of the VL2 peak.  The thick
  solid black curve shows the interior overdensity profile of the
  Lagrangian volume at $z=104.3$, linearly evolved to redshift $z=0$.
  Lagrangian radius is measured relative to the Lagrangian centroid of
  particles in the halo's main progenitor at $z=17.88$.  The thick
  dashed black curve shows the average profile about the Lagrangian
  centroid of the particles inside $r_{200}$ at $z=0$.  For
  comparison, the upper (blue) line and shaded region indicate the
  mean and dispersion in the highest subpeak profile expected from
  supremum statistics (see \S\ref{sec:stats}), while the lower (gray)
  line and shaded region depicts the mean and dispersion expected for
  the peak profile neglecting subpeaks (BBKS).
  \label{fig:delta}
}
\end{figure}

The first ingredient needed is the linear density profile of the
initial peak that collapses to form this halo.  This is plotted as the
solid black curve in Figure \ref{fig:delta}. Given this peak profile,
we predict each shell's invariant $r\times M$ using the spherical
collapse model.  At early times, prior to shell crossing, the mass
enclosed within each shell is a constant, 
$M_{\rm L}=(4\pi/3){\bar\rho}\rL^3$, and the spherical collapse model
predicts the shell's turnaround radius to be
$r_{\rm ta}=0.6\rL/{\bar\delta}(\rL)$.
We assume that each shell's adiabatic invariant is equal to 
$r_{\rm ta}\times M_{\rm L}$.

\begin{figure}
\centerline{\includegraphics[width=0.45\textwidth]{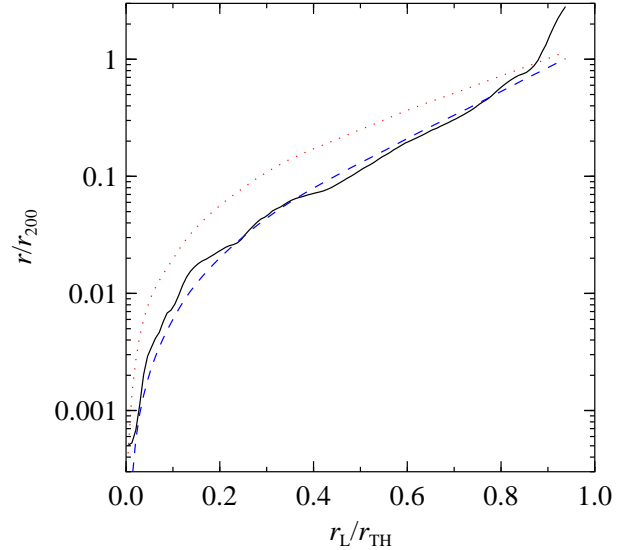}}
\caption{Adiabatic contraction.  The solid black curve shows the
  median radius $r$ at $z=0$ for particles in narrow Lagrangian
  shells, as a function of Lagrangian radius \rL.  The dashed blue
  curve shows the predicted final radius for each shell, under the
  assumption that each shell adiabatically contracts, with adiabatic
  invariants determined from the linear density profile of the initial
  peak at $z=104.3$.  For comparison, 
  the dotted red curve shows the predicted final radius in the frozen
  model, in which $r\propto r_{\rm ta}$ with no subsequent adiabatic
  contraction.
  \label{fig:rfinal}
}
\end{figure}

Given each shell's adiabatic invariant, we can compute the shell's
average final radius within the halo at $z=0$, given the halo's mass
profile $M(r)$: we simply find the radius where $r\times M(r)$ is
equal to $r_{\rm ta}\times M_{\rm L}$ for each shell.  This is plotted
as the dashed blue curve in Figure \ref{fig:rfinal}.  For comparison,
the solid black curve in Fig.\ \ref{fig:rfinal} shows the median
radius $r$ measured at $z=0$ for each shell.  The level of agreement
between the two curves is striking, especially since neither has any
freedom to be adjusted.

It is important to stress that this test is not circular:
this is a direct check that shells conserve $r\times M(r)$, the
prediction of adiabatic evolution.  For a
given $M(r)$ profile, there are many ways that the Lagrangian shells
could add up to give the total mass.  For example, if the halo had
violently relaxed during formation, such that all the orbital actions
became randomized, then the final orbital actions (and hence orbital
radii) would be unrelated to original Lagrangian location, and we
would expect the median $r$ to be independent of $r_{\rm L}$.  This
clearly does not occur within the VL2 halo.

\begin{figure}
\centerline{\includegraphics[width=0.45\textwidth]{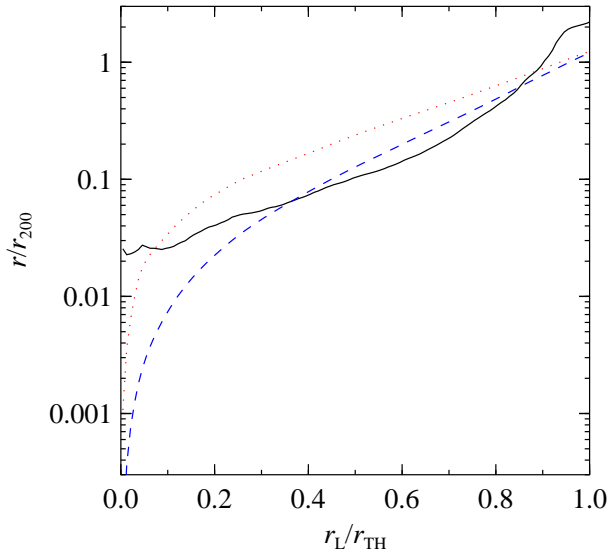}}
\caption{Like figure \ref{fig:rfinal}, but using the $z=0$ Lagrangian
  centroid.  The disagreement at small $r$ shows that the mass at
  small radius, $r\approx 0$, does not originate near the centroid of
  the overall Lagrangian volume, but instead originates near the
  subpeak that was used in Fig.\ \ref{fig:rfinal}.
  \label{fig:badrfinal}
}
\end{figure}

To illustrate that this test could have failed, we plot in Figure
\ref{fig:badrfinal} how this comparison would have looked if we had
used a different Lagrangian centroid.  The ${\bar\delta}(\rL)$ profile
plotted in Fig.\ \ref{fig:delta} was measured relative to the
Lagrangian centroid of the progenitor halo at high redshift,
$z=17.88$.  If we instead use the Lagrangian centroid of all the
particles in the halo at $z=0$, we obtain the result shown in
Fig.\ \ref{fig:badrfinal}.   Here, the agreement between the predicted
and measured radii at $z=0$ is reasonable at large radii, but much
worse near $r=0$.  There are two reasons for
this.  First, the linear density profile ${\bar\delta}(\rL)$ changes
when we measure Lagrangian radii relative to a different centroid, as
illustrated by the difference between the solid and dashed curves in
Fig.\ \ref{fig:delta}.  More importantly, however, the particles
originating near the $z=0$ Lagrangian centroid are not the particles
that are found near $r=0$ in the final collapsed halo.  Rather, the
mass at small $r$ in the final halo largely originated near a sub-peak
within the total Lagrangian volume, that gave rise to the main
progenitor halo at high redshift.  This sub-peak is off-center within
the overall Lagrangian volume (see Fig.\ \ref{fig:delta}), 
but the mass within this off-center peak eventually falls towards
$r=0$ through processes like dynamical friction.  Earlier
work \citep[e.g.][]{Diemand05,Wang10} has already shown that the
mass at small radius within halos typically collapses at high
redshift.  Our result is potentially much more powerful: we now have a
means of quantitatively calculating where material will occur within
the halo, as a function of time.

The level of agreement shown in Fig.\ \ref{fig:rfinal}
is even more remarkable when we consider how simplistic
the prediction is.  To reiterate, we take the spherically averaged
linear density profile ${\bar\delta}(\rL)$, apply the spherical
collapse model to predict each shell's turnaround radius $r_{\rm ta}$,
and then assume that the product $r_{\rm ta}\times M_{\rm L}$ is an
adiabatic invariant.  The formation of the VL2 halo is highly
nonspherical\footnote{The nonspherical collapse of this halo may be
  seen in the movie 
  {\tt http://astro.berkeley.edu/$\sim$mqk/VL2/movie\_10M\_withfly.mp4}},
as is typical in CDM cosmologies, so one potential area
of future work could be to explore whether the adiabatic invariants
may be predicted more precisely by accounting for triaxiality.

\begin{figure}
\centerline{\includegraphics[width=0.46\textwidth]{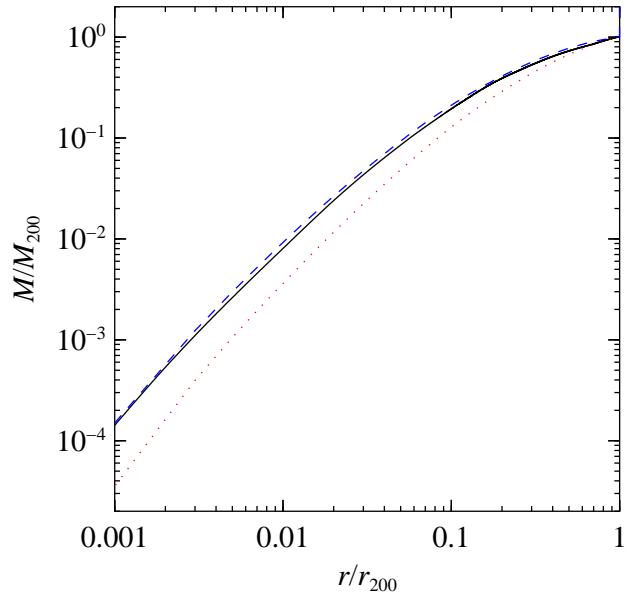}}
\caption{Halo mass profile.  The solid black curve is $M(r)$ measured
  from VL2, the dashed blue curve is our $\rho^{1/2}$ toy model 
  Eqn.\ (\ref{toy2}), and the dotted red curve is the minimal
  contraction model Eqn.\ (\ref{eqn:min}).
  \label{fig:massprofile}
}
\end{figure}

\begin{figure}
\centerline{\includegraphics[width=0.46\textwidth]{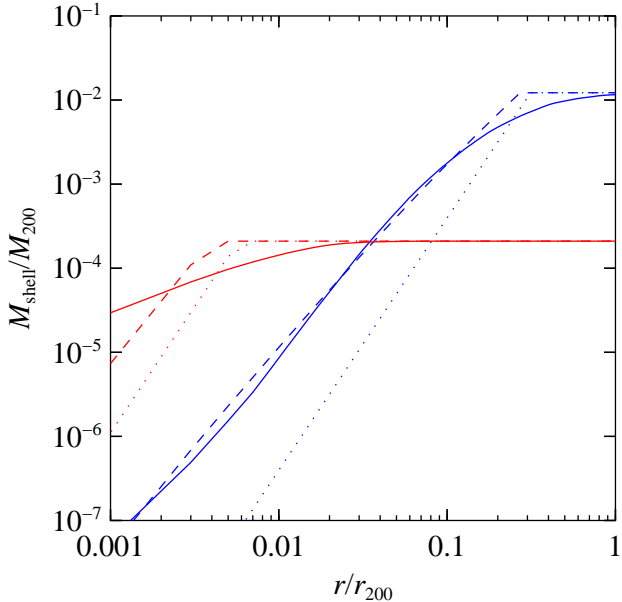}}
\caption{Mass profiles for two particular Lagrangian shells.  The
  solid curves show the profiles measured from VL2, the dotted
  curves are the minimal contraction toy model, and the dashed curves
  are the $\rho^{1/2}$ toy model.  Neither toy model captures the
  shell profiles well, especially at small radius, however the
  $\rho^{1/2}$ model is a close enough approximation to reproduce the
  overall mass profile (see Fig.\ \ref{fig:massprofile}).
  \label{fig:shells}
}
\end{figure}

This simplistic, spherical approach is already adequate to make
considerable progress towards understanding and predicting the mass
distribution within the collapsed halo.  If we know how to calculate
the typical location of each shell within the halo, then we need only
add a prescription for the individual shell profiles to predict the
total mass profile.  In \S\ref{sec:collapse}, we described two
examples of toy models for the profiles of shells: a minimal
contraction model, in which each shell deposits density as a step
function, $\rho_{\rm shell}\propto \Theta(r_{\rm apo}-r)$, and a
slightly more elaborate model in which each shell deposits mass with a
profile $\rho_{\rm shell}\propto \rho_{\rm tot}^{1/2}\Theta(r_{\rm apo}-r)$.
These are both obviously very crude treatments and are no replacement
for a detailed understanding of the orbital structure within halos; we
employ them here simply because they allow us to solve for the total
mass distribution using ordinary differential equations.  We plot the
predictions of these toy models in Figures \ref{fig:massprofile} and
\ref{fig:shells}.  Note 
that, unlike the plots in Figs.\ \ref{fig:rfinal} and
\ref{fig:badrfinal}, this calculation does not use the measured $M(r)$
profile from the VL2 halo at $z=0$.  Rather, these toy models solve
self-consistently for the total mass distribution, placing each shell
at the required radius $r$ given each shell's adiabatic invariant and
the mass profile obtained by summing over all shells.
Fig.\ \ref{fig:shells} shows that this toy model is only a very crude
approximation to the individual shell profiles, but
Fig.\ \ref{fig:massprofile} illustrates that this is already adequate
to predict the halo's radial mass distribution reasonably well.

Before closing this section, we stress that VL2 is just one
halo.  The remarkable agreement between our model predictions and this
simulation's results could, in principle, be a fluke, if (for example)
this halo had an atypical formation history.  Further study of
additional examples \citep[e.g.][]{GHalo,Aquarius}
would be useful to verify the generality of our results.

\section{Peaks and sub-peaks of Gaussian random fields}
\label{sec:stats}

In the previous section, we showed that the radial structure of the
VL2 halo may be understood quite simply.  Lagrangian shells
evolve over time in a manner that, on average, conserves their radial
actions.  These adiabatic invariants may be predicted in a simple
way by applying the spherical collapse model to the spherically
averaged profile of the peak that collapses to form this halo.  Given
this result, even a crude treatment of the mass profiles deposited by
Lagrangian shells suffices to describe the overall mass profile within
the halo reasonably well.
Consistent with previous work, we also found evidence for processes
like dynamical friction that can transport material within
early-collapsing subregions towards the halo center.

Given this success in understanding one particular halo, what can we
say about halo structure more generally?  Our key result is that the
profiles of collapsed halos may easily be understood in terms of the
properties of their precursor peaks.  For most cosmologies of
interest, the properties of the initial peaks are well described by
Gaussian statistics \citep{BBKS}.  

In our model, the radial profile $\rho(r)$ within the halo is
determined by the linear overdensity profile ${\bar\delta}(\rL)$ of
the initial peak.  It is straightforward to determine the statistics
of the interior ${\bar\delta}(\rL)$ profile for Gaussian random
fields.  For example, suppose that we have a peak on scale 
$r_{\rm pk}$ that collapses to make a halo of mass 
$M\simeq (4\pi/3){\bar\rho}r_{\rm pk}^3$, and suppose that the peak
has height ${\bar\delta}(r_{\rm pk})=\delta_{\rm pk}$ and derivative 
$d{\bar\delta}/d\rL(r_{\rm pk})=\delta_{\rm pk}^\prime$ on this scale.
We would like to compute the conditional probability distribution for
the density at interior radii given the peak height and slope at the
outer scale, $P({\bar\delta}(\rL)|\delta_{\rm pk},\delta_{\rm pk}^\prime)$.
In general, the conditional probability distribution $P(X|Y)$ for
Gaussian variables $X$ and $Y$ (with zero mean) is also a Gaussian,
with mean and variance 
\begin{eqnarray}
\langle X|Y\rangle &=& \langle XY\rangle 
\langle YY\rangle^{-1} Y \label{cond_mean} \\
\sigma_{X|Y}^2 &=&  \langle XX\rangle - \langle XY\rangle 
\langle YY\rangle^{-1}  \langle YX\rangle.
\end{eqnarray}
In our case, $X$ would correspond to the small-scale density 
${\bar\delta}(\rL)$, and $Y$ would correspond to the density
$\delta_{\rm pk}$ and slope $\delta_{\rm pk}^\prime$ on scale 
$r_{\rm pk}$.
Given a power spectrum, the required covariances are easy to compute.
In Fig.\ \ref{fig:delta}, the dotted line and gray dashed band
illustrate the mean and dispersion of the $\bar\delta(\rL)$ profile
when we condition on the value and slope of $\bar\delta$ at the
top-hat radius.  The mean profile quickly plateaus inside the top-hat
radius $\rL<r_{\rm TH}$, where $M_{200}=(4\pi/3){\bar\rho}r_{\rm TH}^3$.
In addition, there is considerable scatter about the mean
profile, increasing towards smaller radii.  If we use the $z=0$
Lagrangian centroid, the measured ${\bar\delta}$ profile (dashed line
in the Figure) is quite consistent with this BBKS prediction.
Previous workers have made similar arguments for the expected shape of
the Gaussian peaks that form halos.  \citet{Hoffman85} assumed that
peak profiles follow the unsmoothed matter correlation function, which
BBKS later showed is not typical for Gaussian peaks.
\citet{RydenGunn87} and \citet{Ryden88} used the average BBKS profile,
however they used a smoothing scale much smaller than the top-hat
scale of the halos, which gives a profile extremely atypical for the
peaks that collapse into halos.  More recently, \citet{DelPopolo09}
employed the mean BBKS profile to describe the typical profiles of
Gaussian peaks.

As we have seen in \S\ref{sec:VL}, however, the linear profile
centered on the overall Lagrangian centroid may not be relevant for
computing the structure of the collapsed halo, because processes like
dynamical friction can drag off-center subpeaks towards the halo
center.  We found good agreement between the measured and predicted
adiabatic invariants if we instead used the Lagrangian position of the
sub-peak that formed the earliest halo progenitor at $z=17.88$.  As
Fig.\ \ref{fig:delta} shows, the profile centered on this subpeak is
quite different than the mean BBKS profile used in earlier works.  In
general, we would expect similar behavior in other halos as well.  The
hierarchy of peaks within peaks expected for CDM cosmologies
significantly modifies the structure that we would naively calculate
using the mean peak profile.

One simple way to account for this effect is simply to graft the
profile of the highest subpeak onto the overall peak profile.  So if we
wish to compute the density ${\bar\delta}(\rL)$ for some \rL\ that is
much smaller than the peak size, instead of using the value of 
${\bar\delta}(\rL)$ centered on the halo's Lagrangian centroid, we
instead find the largest ${\bar\delta}$ for all the sub-volumes of
size $r_{\rm sub}=\rL$ within the halo's Lagrangian volume.  This is
effectively what we have done in Fig.\ \ref{fig:delta} by centering on
the $z=17.88$ sub-peak.  Of course, this sub-peak will have
subsubpeaks inside of it, but with a hierarchy of such grafts, we can
construct the effective peak profile that accounts for effects like
dynamical friction.

\begin{figure}
\centerline{\includegraphics[width=0.45\textwidth]{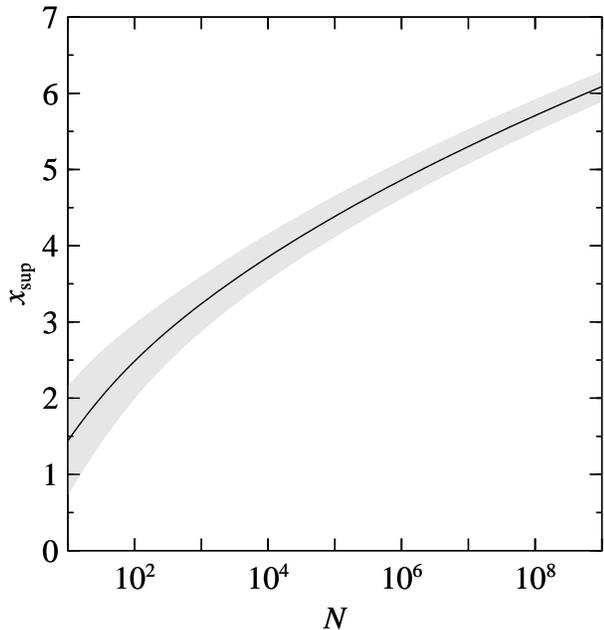}}
\caption{Mean and dispersion of the supremum, $x_{\rm sup}$, the
  largest value of $N$ random Gaussian numbers with zero mean and unit
  variance.  
  \label{fig:supremum}
}
\end{figure}

The statistics of the highest subpeaks are elementary to compute, as
we illustrate with a simple example.  Suppose that $x$ is a Gaussian
random variable with zero mean and unit variance, of which we have $N$
independent samples.  The probability that $y$ exceeds any one of
these $x$ samples is
\begin{equation}
P_1(y) = \int_{-\infty}^y \frac{dP}{dx}dx = 
1-\frac{1}{2}{\rm erfc}\left(\frac{y}{\sqrt{2}}\right),
\end{equation}
and so the probability that $y$ exceeds all $N$ of the $x$ values is
\begin{equation}
P_N(y) = [P_1(y)]^N
\approx \exp\left(-N y^{-1} e^{-y^2/2}/\sqrt{2\pi}\right),
\label{eqn:supremum}
\end{equation}
where the last approximation is valid in the limit of large $N$.  
The double-exponential form of the extreme-value
probability is not specific only to our Gaussian
distribution, but in fact arises generically for any parent
distribution that is sufficiently steep \citep{Bhavsar85}. 
The differential probability that $y$ is the supremum of the $N$
samples is $dP_N/dy$.  In Figure \ref{fig:supremum}, we plot the
typical supremum as a function of sample size $N$.  As might be
expected for a Gaussian parent distribution, the largest value grows
roughly like $(\log N)^{1/2}$ in the limit of large $N$.

It is straightforward to apply this expression to our problem of
computing the distribution of the highest subpeaks.  Here, $P_1$ is
simply the BBKS probability described above, and 
$N=(r_{\rm pk}/r_{\rm sub})^3$ is the number of independent samples of
size $r_{\rm sub}$ inside a peak of size $r_{\rm pk}$.  By
construction, we account for the correlations between the density on
scale $r_{\rm sub}$ and the density and derivatives on scale 
$r_{\rm pk}$, but we neglect any additional spatial correlations among
the subpeaks beyond this. 

We caution that, in general, this expression will overestimate the
height of the effective $\delta(\rL)$ profile, and will underestimate
its scatter.  This is because dynamical friction is not always
effective, especially at low subhalo mass -- which is why so much
substructure persists in CDM halos \citep[e.g.][]{ViaLactea2}.  We are
implicitly assuming that the highest subpeak of size $r_{\rm sub}$
will always find its way to the origin, but for 
$M_{\rm sub} \ll M_{\rm halo}$, the dynamical friction timescale may
exceed the halo's lifetime, meaning that this sub-peak would have to
be carried to $r=0$ by some larger scale structure.  Clearly, it is
not always the case that, for example, the highest peak on scale
$r=10^{-3}$ arises inside of the highest peak on scale $r=10^{-2}$.  A
more careful treatment would be considerably more complicated,
however, so we have opted to make this simplifying assumption.

Bearing this caveat in mind, the blue curve and shaded area in
Fig.\ \ref{fig:delta} plots the mean value and dispersion of the
highest sub-peak, calculated using supremum statistics.  As with the
BBKS curve, we have conditioned the probability on the value and
derivative of the linear overdensity ${\bar\delta}$ on the top-hat
scale corresponding to the measured mass $M_{200}$.  The agreement
between the predicted curve and the actual measured peak profile over
decades in radius is striking.

Another remarkable feature is that the scatter in the effective
profile is much smaller than the scatter in the BBKS profile.  This
occurs because the the dispersion in the largest value of a sample
is considerably smaller than the dispersion of the sample as a
whole, in the limit of large $N$.
Note that this level of dispersion is not indicative of
the full scatter in peak profiles for halos of a fixed mass.  In this
example, we have constrained the profiles to match the value and slope
of the VL2 peak profile at the largest plotted radius.  Other
halos with comparable mass will correspond to peaks with different
heights and slopes at the boundary, and hence their internal profiles
will show different behavior.  

\begin{figure}
\centerline{\includegraphics[width=0.45\textwidth]{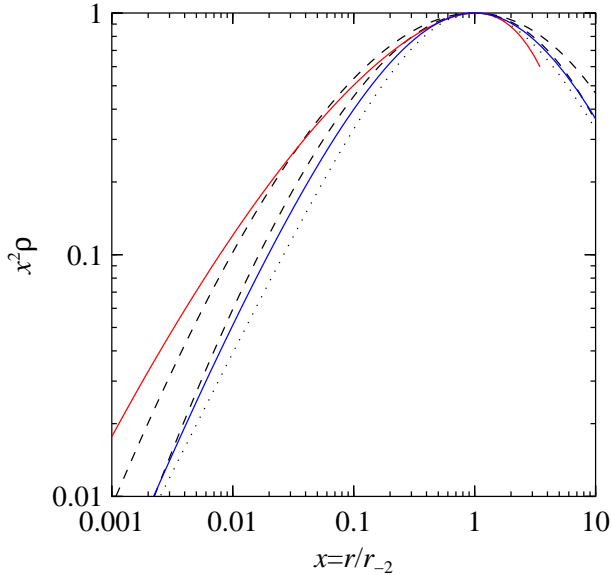}}
\caption{Density profile for peaks with the same height as the VL2
  peak, but with different outer slopes (solid red and blue curves),
  using the $\rho^{1/2}$ toy model.  For comparison, the dotted black
  curve shows the NFW profile, while the two dashed black curves are
  the Einasto profile for $\alpha=0.13$ and 0.17, the range of the
  Aquarius halos \citep{Aquarius}.
  \label{fig:range}
}
\end{figure}

To illustrate this, we plot in Fig.\ \ref{fig:range} the predicted
profiles for initial peaks that have outer slopes different than the
VL2 peak.  The blue and red solid curves show the predicted
halo profiles for peaks with outer slopes either $2\times$ larger, or
$3\times$ smaller, than the VL2 peak.  For comparison, the
dotted curve shows the NFW profile,
\begin{equation}
\rho_{\rm NFW}\left(x=\frac{r}{r_s}\right)=\frac{\rho_s}{x(1+x)^2},
\end{equation}
and the dashed curves show the Einasto profile \citep{Merritt05},
\begin{equation}
\rho_{\rm Ein}\left(x=\frac{r}{r_{-2}}\right)=
\rho_{-2}\exp\left[-2\alpha^{-1}(x^\alpha-1\right)]
\end{equation}
for $\alpha=0.13$ and 0.17, covering the range found in the Aquarius
simulations \citep{Aquarius}.  As the figure shows, varying the outer
profile slope of the initial peak over a reasonable range spans the
profile shapes found in high resolution simulations.

\section{Concentrations}
\label{sec:conc}

The previous sections described a simple model for the structure of
cosmological halos, combining Gaussian statistics with prescriptions
for physical effects like dynamical friction and adiabatic
contraction.  Such a model can have many obvious applications.  We
illustrate with one example in this section.

In our model, the shape of the final collapsed halo profile is simply
related to the shape of the linear density profile of the original
peak.  One of the important parameters used to describe halo profiles
is the concentration, which we can define as the ratio between the
halo's radius $r_{200}$, and $r_{-2}$, the radius where the local density slope
is $d\log\rho/d\log r = -2$.  Many papers have attempted to quantify
the typical halo concentration as a function of mass 
\citep[e.g.][and references therein]{Munoz10}.  Using our model, we
can try to predict this $c(M)$ relation.

The halo concentration basically measures how quickly the halo slope
rolls over from near -3 to -2, and in our model this is controlled by
the outer height and slope of the initial peak (mainly the latter).
The distribution of the heights of the peaks producing halos is set by
the distribution of collapse thresholds.  \citet{Bond96} proposed a
simple ellipsoidal collapse model to predict the distribution of
collapse thresholds, and \citet{Sheth01} found that the predictions of
the \citeauthor{Bond96} model are reasonably approximated by the
fitting function
\begin{equation}
\delta_{\rm ec}=\delta_{\rm c} \left[1 + \beta
\left(\frac{\sigma}{\delta_{\rm c}}\right)^{2\gamma}\right],
\end{equation}
where $\delta_{\rm c}=1.686$ is the spherical collapse threshold,
$\sigma^2(M)$ is the variance of linear density fluctuations smoothed
on mass scale $M$, and the fitting function parameters are
$\beta=0.47$ and $\gamma=0.615$.  
\citet{Robertson09} showed that this expression is in broad agreement
with the heights of peaks that produce halos in $\Lambda$CDM
simulations, so we will adopt it here.  More importantly, we also need
the typical outer slopes of the same peaks.  This has been less well
quantified in the literature.  For high mass halos, with 
$\sigma(M)\ll \delta_{\rm c}$, we can use simple Gaussian statistics
to predict the outer slopes 
${\bar\delta}^\prime\equiv d{\bar\delta}/d\log\rL$.  The typical
slope for a peak of height ${\bar\delta}$ is
\begin{equation}
\langle{\bar\delta}^\prime|{\bar\delta}\rangle=
\int_{-\infty}^0 P({\bar\delta}^\prime|{\bar\delta}) d{\bar\delta}^\prime,
\label{eqn:slopebbks}
\end{equation}
where $P$ is the conditional Gaussian probability distribution.
Note that this expression is different than Eqn.\ (\ref{cond_mean}),
since we require that the slope ${\bar\delta}^\prime<0$, since
otherwise this peak would not collapse on mass scale $M$ but instead
some larger mass \citep{BBKS}.  
This expression works well for high mass halos,
$M\gg M_\star$, where $M_\star$ is the characteristic nonlinear mass
scale satisfying $\sigma(M_\star)=\delta_{\rm c}$.  However, this
significantly underestimates the magnitude of the outer slopes for low
mass halos, $M\lesssim M_\star$, which are considerably steeper than
Eqn.\ (\ref{eqn:slopebbks}) predicts.  This appears to occur for the
same reason that these low-mass halos are anti-biased: such halos tend
to avoid high-density regions.  More precisely, the peaks that produce
low mass halos tend to occur within underdense environments,
presumably because similar peaks inside of overdense regions do not
lead to low mass halos, but rather are incorporated into higher mass
halos.  This is the reason why the average peak profile plotted as the
dotted curve in Fig.\ \ref{fig:frozen} changes sign.  As the figure
shows, the outer profiles for such halos are extremely steep, which we
argued in \S\ref{sec:frozen} was the origin of the high
concentrations of these halos. 

Given the lack of a theory to describe this effect, we have attempted 
to measure the typical outer slopes of initial peaks.  We performed a
low-resolution $\Lambda$CDM simulation for a volume 256 $h^{-1}$ Mpc
on a side, identified halos at $z=0$ down to mass 
$M\approx 10^{12}h^{-1}M_\odot$, and stacked the linear density
profiles of the halos' precursor peaks in narrow mass bins.  Very
crudely, we found that a rough scaling 
\begin{equation}
\frac{d{\bar\delta}}{d\log\rL} \approx -1.5 -\log(1+\sigma^2)
\label{slopefit}
\end{equation}
reasonably described the mass range that we measured.  

\begin{figure}
\centerline{\includegraphics[width=0.45\textwidth]{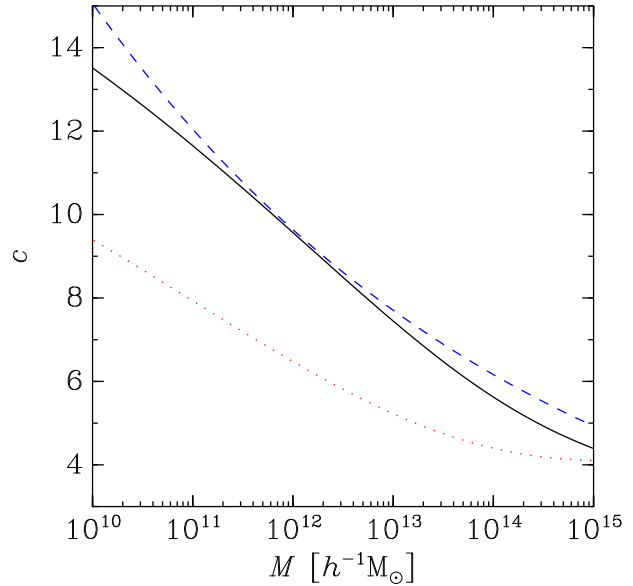}}
\caption{Mean concentration relation $c(M)$.  The solid black curve
  shows our model (see text), while the dotted red curve shows the
  prediction from using the naive peak slopes from Gaussian
  statistics.  For comparison, the dashed blue curve shows the power
  law $c\propto M^{-0.097}$ found in N-body simulations \citep{Munoz10}.
  \label{fig:conc}
}
\end{figure}

Given this expression, and our expression for the average peak height
as a function of mass, we use Eqn.\ (\ref{eqn:supremum}) to predict
the average peak profile, and then Eqn.\ (\ref{toy2}) to predict the
halo profile, from which we measure the concentration.  Figure
\ref{fig:conc} shows the result.  The solid black curve in the figure
is our prediction, while the dashed blue curve shows the scaling
$c\propto M^{-0.097}$, as found in cosmological simulations
\citep{Munoz10}.  For comparison, the red dotted curve shows how the
concentration would depend on mass if the outer slopes followed simple
Gaussian statistics.  The agreement between our prediction and the
results of simulations is, once again, quite good.  At low masses,
this model appears to underpredict the concentrations, but this is
likely because Eqn.\ (\ref{slopefit}) underestimates the slopes of
halos below $M<10^{12}M_\odot$.  At high masses, note that our
concentrations appear to saturate near values $c\approx 4$.  This
occurs because, at high masses, peaks no longer are anti-biased, and
hence their outer slopes are given by the shallow values predicted
from Gaussian statistics.

\section{Conclusions}
\label{sec:conclude}

In this paper, we have presented a model to explain the origin
of the nearly universal density profiles of dark matter halos found in
N-body simulations.  We argued that the physics setting the shape of
halo radial profiles is extremely simple.  We find that adiabatic
contraction sets the basic shape of the halo profile, and that the
conserved quantities in this contraction, i.e.\ the adiabatic
invariants, are determined by the linear density profile of the
initial peak.  We further argue
that, because of dynamical friction, the hierarchy of peaks within
peaks significantly alters and {\em simplifies} the effective peak
profile setting the adiabatic invariants and hence the halo profile.  

We have compared our model predictions to N-body simulations, and
found striking agreement.  In particular, the detailed mass
distribution of the high-resolution Via Lactea-II halo is quite
consistent with our model, and provides strong evidence for the
importance of both adiabatic contraction and dynamical friction.  We
then showed how this model may be used to predict the statistics of
halo properties, focusing on the example of the mean concentration
relation $c(M)$. 

Our model, if correct, could have additional implications beyond what
we have discussed so far.  One example that has attracted considerable
attention is the asymptotic inner profile of dark matter halos, which
can be important for dark matter annihilation 
\citep[e.g.][]{Bergstrom98,Kuhlen09,Kuhlen10,Reed10}.
For CDM-like power spectra, our models do not produce power-law
central cusps, but instead diverge logarithmically.  The local slope
$d\log\rho/d\log r$ rolls over very slowly with radius, asymptotically
approaching zero slope at $r=0$.  As we noted above, recent high
resolution simulations have found qualitatively similar behavior,
where the halo slope $d\log\rho/d\log r$ rolls smoothly down to the
resolution limits of the simulations.  As we showed in Figure
\ref{fig:range}, the Einasto profiles used to fit these halos are quite
similar to the predictions of our model.

As we have emphasized in this paper, our model is not complete,
because it lacks a detailed understanding of the mass profiles
deposited by Lagrangian shells.  For simplicity, we have adopted
a simple ansatz that is loosely motivated by our previous study of
self-similar triaxial collapse (Paper I), however in the future we
intend to examine the orbital distribution within halos in more detail.
Our study of self-similar collapse has already shown that the
evolution of orbits in time-varying triaxial potentials is quite
interesting and can often be surprising, for example in the different
ways that box orbits and loop orbits respond to changes in the
potential.  This is the subject of ongoing work.

Lastly, we note that it would be useful to extend the comparison of
our models to other high resolution halo simulations.  In particular,
it would be worthwhile to explore the limits of this model, which
so far appears to work surprisingly well in regimes (e.g.\ $r\sim
r_{\rm vir}$) where we might naively have expected it to fail.  One
regime where this model is likely to break down is the case of nearly
1:1 major mergers in halos, in which both progenitors have comparable
central density and both contribute to the central core.  It would be
interesting to see whether we can find adiabatic invariants to
describe the dynamics even in such extreme cases.

\acknowledgments{We thank Peter Coles and Andrey Kravtsov for useful
  discussions.  We also thank Dmytro Iakubovskyi and Barbara Ryden for
  bringing several relevant publications to our attention.
}


\end{document}